\documentclass[twocolumn]{aastex62}

\usepackage{soul}
\usepackage[graphicx]{realboxes}
\usepackage{rotating,tabularx}

 \usepackage{mathtools}
\usepackage[]{algorithm}
\usepackage{algpseudocode}
\usepackage{ifthen}
\usepackage{amssymb}
\usepackage{amsmath}
\usepackage[]{algorithm}
\usepackage{algpseudocode} 
\usepackage{fontawesome}

\newcommand{\msun}{\ensuremath{\, M_\odot}} 
\newcommand{\kpc}{\ensuremath{\, {\rm kpc}}}

\newcommand{\boldSigma}{\boldsymbol{\Sigma}}

\newcommand\boldrho{\boldsymbol{\rho}}

\newcommand{\boldcorr}{\boldsymbol{R}}

\newcommand{\Ngal}{N_{\rm gal}}

\newcommand{\rhog}{\rho_g} 
\newcommand{\rhos}{\rho_{\star}}
\newcommand{\rhom}{\rho_m}

\newcommand{\fg}{f_g}   

\newcommand{\dg}{\delta_g}   
\newcommand{\ds}{\delta_{\star}}
\newcommand{\dm}{\delta_m}

\newcommand{\dfg}{\delta_{f,g}}   
\newcommand{\dfs}{\delta_{f,\star}}

\newcommand{\rgs}{r_{g\star}}  
\newcommand{\rgm}{r_{gm}} 
\newcommand{\rsm}{r_{\star m}}

\submitjournal{ApJ}

\shorttitle{Correlated DM, Gas, and Stellar Profiles}
\shortauthors{A. Farahi et al.}

\begin{document}

\title{Correlations of Dark Matter, Gas and Stellar Profiles in Dark Matter Halos}

\correspondingauthor{Arya Farahi}\email{arya.farahi@austin.utexas.edu}

\author[0000-0003-0777-4618]{Arya~Farahi}
\affil{Department of Statistics and
Data Science, The University of Texas at Austin, TX 78712, USA}
\author[0000-0002-6766-5942]{Daisuke~Nagai}
\affil{Department of Physics, Yale University, New Haven, CT 06520, USA}
\author[0000-0003-3312-909X]{Dhayaa~Anbajagane}
\affil{Department of Astronomy and Astrophysics, University of Chicago, Chicago, IL 60637, USA}
\affil{Kavli Institute for Cosmological Physics, University of Chicago, Chicago, IL 60637, USA}

\begin{abstract}
Halos of similar mass and redshift exhibit a large degree of variability in their differential properties, such as dark matter, hot gas, and stellar mass density profiles. This variability is an indicator of diversity in the formation history of these dark matter halos that is reflected in the coupling of scatters about the mean relations. In this work, we show that the strength of this coupling depends on the scale at which halo profiles are measured. By analyzing the outputs of the IllustrisTNG hydrodynamical cosmological simulations we report the radial- and mass-dependent couplings between the dark matter, hot gas, and stellar mass radial density profiles utilizing the population diversity in dark matter halos. We find that for the same mass halos the scatters in the density of baryons and dark matter are strongly coupled at large scales ($r>R_{200}$); but the coupling between gas and dark matter density profiles fades near the core of halos ($r < 0.3 R_{200}$). We then show that the correlation between halo profile and integrated quantities induces a radius-dependent additive bias in the profile observables of halos when halos are selected on properties other than their mass. We discuss the impact of this effect on cluster abundance and cross-correlations cosmology with multi-wavelength cosmological surveys.
\end{abstract}


\section{Introduction} \label{sec:intro}
The diversity in the internal structure of halos, such as those that host galaxy clusters, is not only a function of the fundamental physics -- nature of dark energy models, the sum of neutrino masses, non-Gaussian primordial fluctuations, the nature of gravity, and astrophysics -- but also dependent on the gas and galaxy formation physics \citep{Voit:2005,Weinberg:2013,Pratt:2019}.
Many large scale multi-wavelength astronomical surveys (such as Rubin, Roman, eROSITA, Simons Observatory, CMB-S4, and CMB-HD) are designed to find and measure the internal properties of a large number of group-size and cluster-size halos across many wavelengths. These upcoming datasets offer unprecedented discovery opportunities regarding the nature of the dark matter-gas-galaxy connection over cosmic time. 

According to the standard model of cosmology, the gravitational amplification of the primordial matter density field leads to the collapse and formation of dark matter halos \citep{Frenk12}.
The concordance $\Lambda$CDM model predicts that halos form initially via the coherent infall of matter and grow in mass via a combination of further infall and hierarchical merging. Gravitational relaxation drives the phase-space structure of dark matter halos of all scales from small galactic satellites to the most massive galaxy clusters \citep{Ludlow:2013}. While it has been shown that cold dark matter halos have a universal density profile \citep*{NFW}, their internal structure (e.g., the size of the dark matter core) is strongly correlated with the formation epoch of halos \citep{Wechsler:2002}.
Another hint of coupling comes from the self-similar evolution models. These models assume that dark matter, gas, and stars are gravitationally coupled in the infall region \citep{Fillmore1984,Bertschinger:1985}, and predict that the dark matter, gas and stellar density of halos are strongly coupled.

The dynamics and evolution of dark matter, gas, and stellar particles within halos are not governed only by gravity. There are also thermal and non-thermal processes, including gas cooling, star formation, and feedback from supernovae (SNe) and active galactic nuclei (AGN), that alter the energy and momentum of the baryonic particles. The dark matter phase-space is indirectly impacted by these processes due to the non-linear gravitational coupling of baryons and dark matter \citep{Anbajagane2021}, a phenomenon sometimes known as baryon back-reaction.
These non-gravitational effects can also decouple collisional and collisionless particles through redistribution of mass, momentum and energy of the collisional gas relative to the collisionless dark matter and stars \citep{Ostriker:2010, Gaspari:2012, Peirani:2017}. The scale at which this decoupling occurs is sensitive to the complex interplay of the structure formation and galaxy formation physics in dark matter halos. Even though the interplay between gravitational and non-gravitational effects is complex, the scales at which the gravitational coupling is suppressed can inform us about the non-gravitational effects.

Motivated by this argument, we study the radial-dependence of the coupling between collisional and collisionless particles by measuring the correlation between scatter of halo density profiles at different radial- and mass-scales.
Suppose $P(\rho_{\rm dm}, \rho_{\star}, \rho_{\rm gas} \mid M, z, r)$ is the joint distribution of dark matter, gas, and stellar density profiles at radii $r$ for halos of mass $M$ at redshift $z$, and let us assume that this probability distribution has a multivariate log-normal form. Under this assumption, $P(\rho_{\rm dm}, \rho_{\star}, \rho_{\rm gas} \mid M, z, r)$ is fully specified by a mean density profile and a covariance matrix. 
We propose to measure the coupling of different phases of matter by measuring the off-diagonal elements of this covariance matrix.  
By analyzing the outputs of IllustrisTNG simulations, we show that the correlation coefficient -- off-diagonal element of the covariance matrix -- between $\rho_{\rm dm}$, $\rho_{\star}$, $\rho_{\rm gas}$ is $>0.5$ and the correlation between collisional and collisionless particles is suppressed at scales of $r < 0.3 R_{200}$.

This coupling has implications for modeling the mass distribution of dark matter halos using cluster observables. Consequently, the covariance between the baryon and dark matter density profiles not only provides new insights on the interactions between dark matter, gas, and stars at different scales, but is also an important model ingredient when performing multi-wavelength cluster mass calibration or a cross-correlation analysis.
For instance, when clusters are selected on their observables, other than halo mass, this covariance matrix can induce additive bias in the expected properties of halos, such as the weak-lensing signal \citep{2022:Zhang}. In this paper, we quantify the impact of this covariance on the expected cumulative mass profile for a optically-selected cluster sample.

The rest of this work is organized as follows. In Section~\ref{sec:sim}, we describe our simulation ensemble and in Section~\ref{sec:method} our population model and the property covariance. We present our main results, the mean and the correlation statistics, in Section~\ref{sec:results}. We discuss the implications of the halo profile covariance in cosmological studies.  in Section~\ref{sec:discussion}. Finally, we conclude in Section~\ref{sec:conclusion}

We use the halo mass definition of $M_{200}$, the mass enclosed within a sphere of $R_{200}$ whose enclosed average overdensity is $200$ times the critical density of the TNG universe at a given epoch. 

\section{Numerical Simulations}
\label{sec:sim}

In this work, we analyze the density profiles of dark matter, gas and stars extracted from the IllustrisTNG simulations \citep[see,][for the details of TNG model implementation]{Nelson:2018, Pillepich:2018, Springel:2018, Marinacci:2018, Naiman:2018}. The simulations' snapshots and catalogs are provided publicly by the IllustrisTNG team \citep[see,][for data release paper]{TNG_data_release}.
The TNG simulation runs are based on the moving-mesh code, Arepo \citep{Springel:2010EMesh, Weinberger:2020} that incorporates radiative cooling, star-formation, feedback from SNe and AGN, and magnetic fields \citep{Weinberger:2017, Pillepich:2018}. It has been shown that these simulations produce a realistic population of galaxy-size \citep{Genel:2018} and cluster-size \citep{Barnes:2018} halos.

The IllustrisTNG sample used in this work consists of a subset of halos with mass $10^{13}\,M_{\odot}$ and above at redshift $z = 0$ from the fiducial TNG100 run. Halos are identified using a ``friends-of-friends'' percolation algorithm \citep{Einasto:1984}. Our observables are the dark matter, gas and stellar differential density profiles. They are computed in spherical annuli with respect to the center of a halo. This center is defined as the minimum of the local gravitational potential which is the most bound dark matter particle identified by the \textsc{Subfind} algorithm \citep{Springel:2001, Dolag:2009}.

The profiles are extracted in $25$ cluster-centric log-spaced radial bins that cover the range $r=(0.05\,$-$\,5.0)\,R_{200}$. To ensure that we include all particles, we first read in the entire simulation snapshot and then select particles in spheres of radius $5.0\,R_{200}$ around every halo of interest, rather than just using the FoF particle set for each halo. For the dark matter, stellar and gas profiles we included all particles or cells, respectively, regardless of whether they are bound to the central halo, bound to the substructure, or are unbound. For the gas profiles, we remove all gas cells that are currently in the wind phase, i.e., those that have recently experienced either a SNe or AGN feedback event and are currently hydrodynamically decoupled.

\section{Analysis Method}
\label{sec:method}

We model the conditional statistics of the log-densities, $\log\boldrho$, as a function of the normalized radial distance from the center of the halo $x =r/R_{200}$ and the halo mass $M_{200}$. $\log\boldrho$, the three-dimensional vector of the log-density profiles $[\log\rho_{\rm dm}, \log\rho_{\rm gas}, \log\rho_{\star}]$, is a random variable modelled with a multivariate normal distribution
\begin{equation}
P(\log\boldrho\,|\, x, M_{200}) \propto \exp\{ - \frac{1}{2}\, \delta\boldrho^\top\, \boldSigma^{-1}\, \delta\boldrho  \}   
\end{equation}
where $\delta\boldrho = \log\boldrho - \langle \log\boldrho\,|\, x, M_{200} \rangle$. The model is fully specified with the average log-density vector $\langle \log\boldrho\,|\, x, M_{200} \rangle$ and the property profile covariance matrix $\boldSigma = {\rm Cov}(\log\boldrho \,|\,x, M_{200})$ at a given radial and halo mass scales. We employ the \cite{Gelman:2015stan} recommended approach to decompose $\boldSigma$ into a correlation matrix $\boldsymbol{R}$ and a scale vector $\boldsymbol{\sigma}$  \citep[see also,][]{Barnard:2000}: 
\begin{equation} \label{eq:sigmaE_decomp}
	\boldSigma = {\rm diag}(\boldsymbol{\sigma}) \, \boldcorr \, {\rm diag}(\boldsymbol{\sigma})
\end{equation}
${\rm diag}(\boldsymbol{\sigma})$ is a diagonal matrix of standard deviations. In our work, we make the assumption that both $\boldsymbol{R}$ and $\boldsymbol{\sigma}$ are a function of halo mass and distance from the center of halo.

In this work, we employ the \textsc{Kernel Localized Linear Regression} \citep[KLLR \href{https://github.com/afarahi/kllr}{\faGithubAlt},][]{Farahi:2022kllr} method to regress the logarithm of the dark matter, gas, and stellar density profiles as a function of $x$ and for TNG halos in a given mass bin.  \textsc{KLLR} performs a kernel-weighted least squares fitting and reports the average profile and the covariance between profiles at fixed $x$ \citep[see Equations (8)--(13) in][]{Farahi:2022kllr}. In empirical settings, where one need to deal with low signal-to-noise (SNR) measurements, population-based inference methods like \textsc{Population Profile Estimator} \citep[PoPE \href{https://github.com/afarahi/PoPE}{\faGithubAlt},][]{Farahi2021Pope}, are more suitable.
\textsc{PoPE} can be used to perform inference both in high SNR and low SNR regimes; however since for high SNR data the statistics derived by \textsc{KLLR} and \textsc{PoPE} are consistent, we refrain from using \textsc{PoPE} for our main analysis here as the method is computationally demanding for large sample sizes such as the one we have from simulations.

\section{Results}
\label{sec:results}

\begin{figure}
     \centering
     \includegraphics[width=0.45\textwidth]{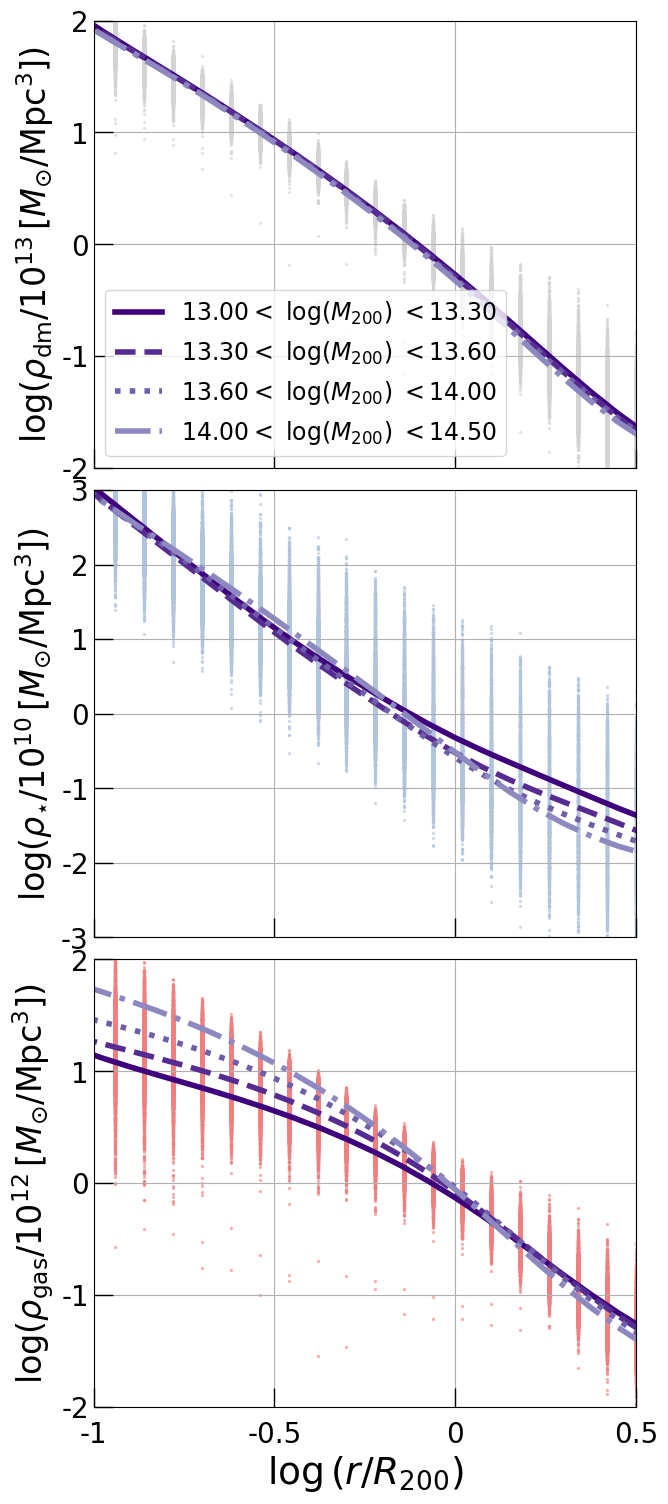} 
\caption{The average log-density profile of dark matter (top panel), stars (middle panel), and gas (bottom panel) for TNG halos at redshift zero. The data points are log-density profiles per halo and and the lines are average profiles per mass bin.}
     \label{fig:TNG_mean_profile_z0}
\end{figure}

Figure~\ref{fig:TNG_mean_profile_z0} shows the dark matter (top panel), stellar (middle panel), and gas (bottom panel) density profiles as a function of normalized radii from the center of halo. Dark matter shows a universal profile with little mass dependence and small population variance ($15\%$ scatter for $r <R_{200})$. However, there is a significant variability in the gas and stellar density profiles at fixed radial distance. A fraction of this variance can be explained with the mass dependence of the density profiles. For instance, at $r=3 \times R_{200}$, the stellar density of cluster-size halos is $>3$ times larger than that of milky-way size halos. To account for the mass-dependence, we split profiles into narrow mass bins and then use the \textsc{KLLR} method to regress against distance scales.

\subsection{Slope and scatter statistics}

\begin{figure*}
     \centering
     \includegraphics[width=0.33\textwidth]{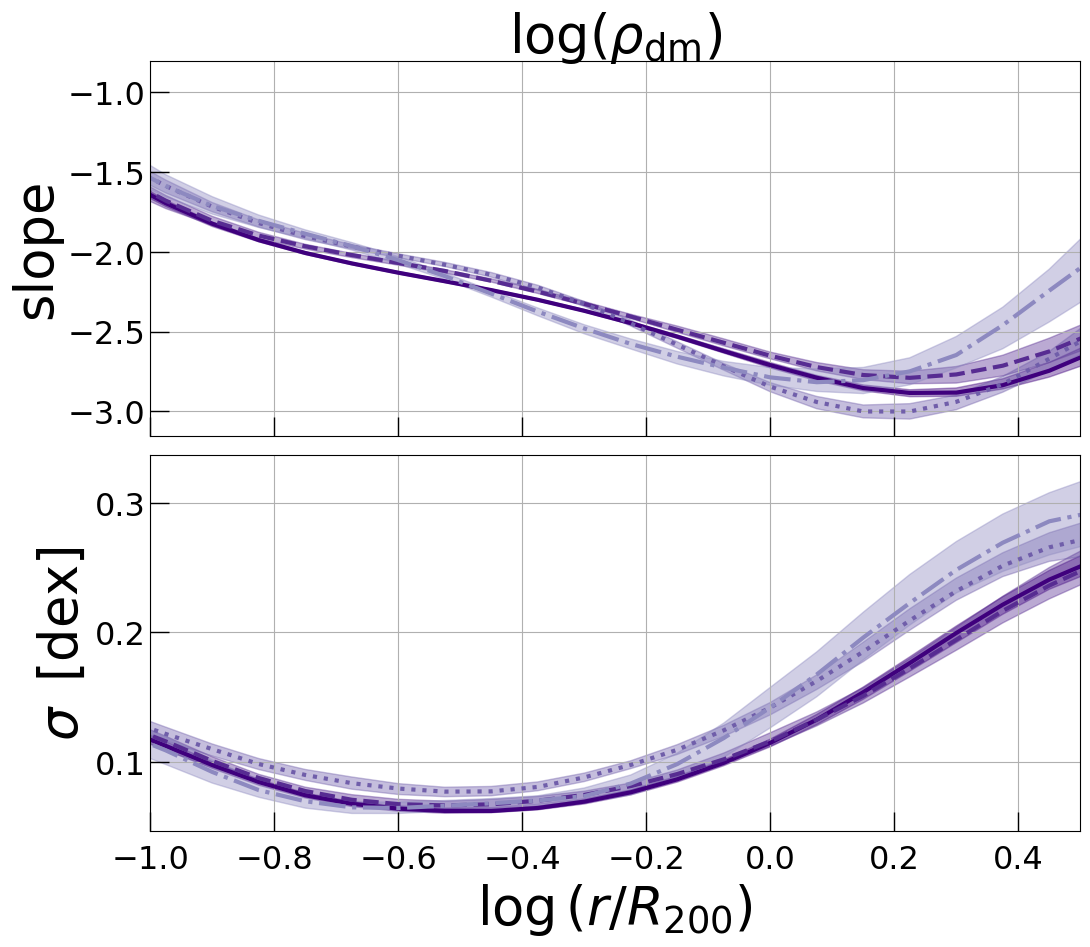} 
     \includegraphics[width=0.31\textwidth]{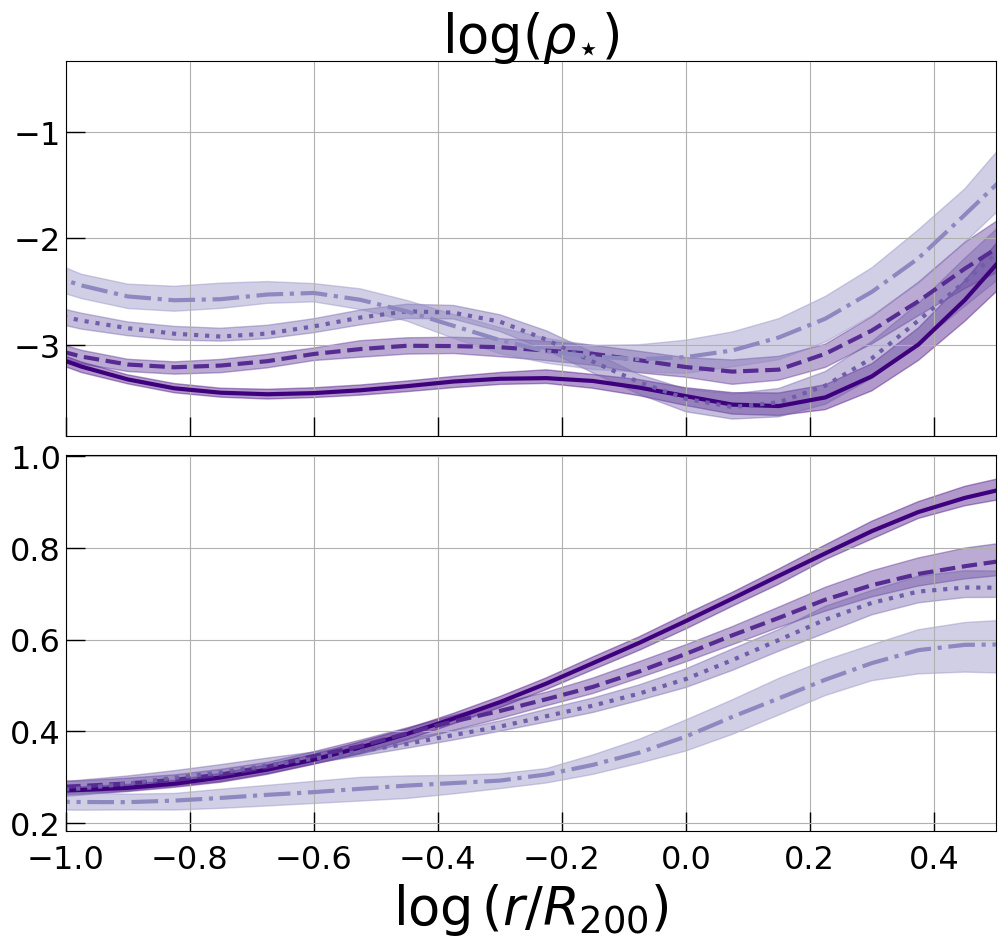} 
     \includegraphics[width=0.31\textwidth]{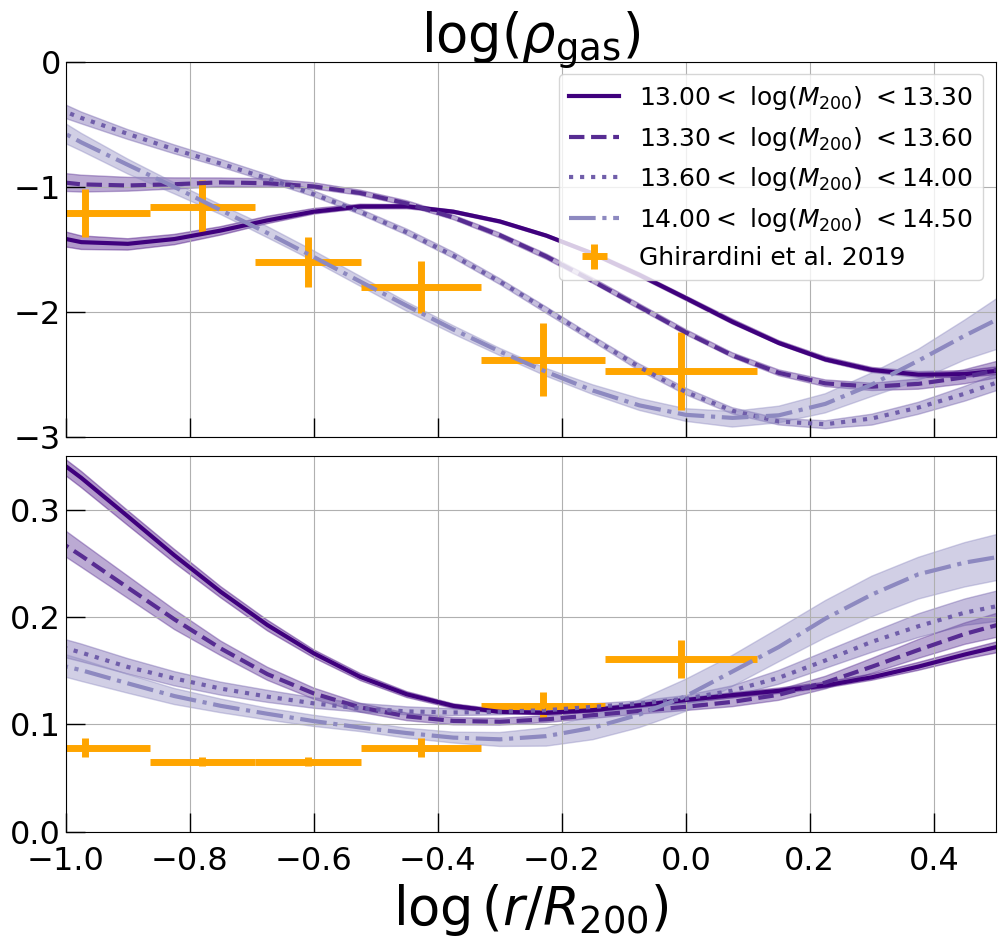} 
\caption{The fit parameters (logarithmic slope and log-scatter) of the average log-density profiles of dark matter (left panel), star (middle panel), and gas (right panel) particles for TNG halos of different mass bins at redshift zero. The orange data points are measurements from the observed sample by \citet{Ghirardini:2019}.} The provided confidence regions are 68\% statistical uncertainty intervals based on 1,000 sample bootstraps.
     \label{fig:TNG_mean_profile_fits_z0}
\end{figure*}

The \textsc{KLLR}-estimated logarithmic slope and log-normal scatter of dark matter, gas, and stellar density profiles are shown in the top and bottom panels of Figure~\ref{fig:TNG_mean_profile_fits_z0}. The logarithmic slope and conditional scatter for the dark matter density profiles are almost independent of the halo mass; but for the gas and stellar density profiles both the slopes and the scatter are mass-dependent. The logarithmic slope of the stellar density profile at fixed halo mass is almost constant and monotonically increases with halo mass, while the slope of the gas density profile is neither constant nor monotonic with halo mass. This implies that a simple linear parameterization of the density profile would fail to model the mass-dependence of the gas density profile. The large scatter in Figure~\ref{fig:TNG_mean_profile_fits_z0} suggests that even after accounting for the mass dependence in the density profiles, there is a significant ($>0.1\,\,[{\rm dex}]$) scatter in gas and stellar density profiles.

The outskirts of the mean dark matter and gas density profiles in massive halos have received significant attention in recent times from both theory and observations \citep[][and references therein]{Walker:2019}, and can serve as probes of astrophysics and of cosmological structure formation \citep[e.g.,][for recent works]{Aung2021ShockSplash, Anbajagane:2021}. In the bottom panels of Figure \ref{fig:TNG_mean_profile_fits_z0}, we see that in the cluster core the dark matter density profile has the least scatter of the three profiles, by about a factor of 2 or more. However, as one moves to the cluster outskirts and towards the infall regime, the \textit{gas} density has the lowest scatter, not the dark matter. Note that this behavior is consistent across the whole halo mass range being probed. Interestingly, at the cluster radius, $r = R_{200}$, both the gas and dark matter density profiles have roughly the same scatter of 0.1 dex.

The orange data points in Figure~\ref{fig:TNG_mean_profile_fits_z0} present the radial profile measurements of the electron density profile of the intracluster medium out to $R_{200}$ by \citet{Ghirardini:2019}, who employed a low-redshift cluster sample with $M_{200} > 6 \times 10^{14}\msun$ selected from the \emph{Planck} all-sky survey.  
While their cluster sample is more massive than the TNG halos, the logarithmic slope of the electron density profile is consistent with that of the most massive halo bin considered in this work. The inferred scatter is also comparable to the estimated scatter using TNG except for $r < 0.25 R_{200}$, where the empirically inferred scatter is smaller than the scatter seen in the simulation.

\subsection{Correlation statistics}

A measurement of the population correlation gives us a thorough picture of the coupling between the density profiles. Figure~\ref{fig:TNG_corr_profile_z0} shows radial-dependence of coupling between the densities of collisionless dark matter, collisionless stars, and the collisional gas at fixed halo mass. Our results suggest that in the infall region where $r > R_{200}$, all three principle phases of matter are strongly coupled.  These coupling coefficients decline in the core of halos; notably the coupling of gas and dark matter becomes suppressed  (particularly for $M_{200} < 4 \times 10^{13}\,[M_{\odot}]$). These trends remain almost intact up-to redshift 2 (see Appendix \ref{app:z_evo_prop_corr}).
For cluster-size halo, the infall region is where gravity  dominates other forces, and the impact of feedback from AGN and SN is essentially non-existent. The infall region is about a few Mpc away from the halo center, while the region of AGN influence is within the few hundreds kpc of the halo core \citep{Kauffmann:2019}. At large scales, the positive correlation between all principle phases of matter reflects this gravitational coupling. 

In the inner region ($r < 0.3\,R_{200}$), the correlation between dark matter and gas densities is suppressed for massive halos with $M_{200} < 4 \times 10^{13}\,M_{\odot}$.  The gas decoupling scale\footnote{We define gas decoupling scale as a scale at which the correlation between dark matter and gas densities fades.} of these halos is $\sim 100\,\kpc$.
There are several factors, including gas cooling, star formation rate, and feedback, that might contribute to this decoupling. We find that the average gas density profile, unlike stellar and dark matter density profiles, of fixed halo mass is sensitive to the star formation rate of the central galaxy, which provides evidence that the gas--dark matter decoupling is sensitive to galaxy formation physics.
We notice that the scale of gas decoupling coincides with the AGN feedback sphere of influence in the TNG simulations \citep{Kauffmann:2019}. This association suggests that strong winds from AGN and SNe evacuate gas from the halo center and are one of the key factors in gas-dark matter decoupling \citep{Terrazas:2020,Pellegrini:2012}. However, a future study should assess the contribution of each one of the aforementioned factors, which can be radial- and mass-dependent, in influencing this decoupling.

Another intriguing observation is that while the gas and dark matter of low mass halos decouples at scales less than $0.3 \times R_{200}$ (Figure~\ref{fig:TNG_corr_profile_z0}, left panel), the stellar density is strongly coupled with dark matter density and decoupled with gas density (Figure~\ref{fig:TNG_corr_profile_z0}, middle and right panels, respectively). The presence of stars, which have a more concentrated density profile, steepens the gravitational potential and causes the whole halo to contract. 
This might suggest that star formation can contribute to the correlation between the different collisionless components.
The collisional particles -- the gas -- respond to feedback and cooling processes, which redistribute their energy and momentum. This consequently impacts the collisionless particles -- the stars and dark matter -- due to the non-linear gravitational coupling between the matter components, and alters the particles' phase-space distribution \citep{Anbajagane2021}. 
The stellar and dark matter density correlation in the core (Figure~\ref{fig:TNG_corr_profile_z0}, middle panel) might also be linked to the correlation with the formation time \citep{Wechsler:2002,DeLucia:2007, GoldenMarx:2018}.

We use ``closed-shell'' (``open-shell'') regime to refer to scales at which the total baryon fraction is (not) conserved. To study the transition between these regimes, we utilize the correlation between hot gas and stellar mass fraction as proposed in \citet{Wu:2015} and \citet{Farahi:2019anti}. We define differential gas (stellar) mass fractions as the gas (stellar) mass divided by the total mass within a radial shell, $f_{\rm gas/\star} = M_{\rm gas/\star}(r)/M_{\rm tot}(r)$.
Figure~\ref{fig:TNG_fb_profile_z0} shows the correlation between differential gas and stellar mass fractions as a function of radial- and halo mass-scale. At radial-scales larger than $R_{200}$, there is anti-correlation between gas and stellar fraction that suggests baryons in the infall radial shells are approximately conserved \citep{Wu:2015, Farahi:2018, Farahi:2019anti}. But in the core of low-mass halos with $M_{200}<10^{14}\,\msun$, the anti-correlation between $f_{\rm gas}$ and $f_{\rm \star}$ vanishes (Figure~\ref{fig:TNG_fb_profile_z0}). This change in correlation indicates the transition from closed-shell to open-shell regime.
The transition trend from closed-shell to open-shell regime is consistent with the correlation trend that is observed in density profiles of gas and dark matter (Figure~\ref{fig:TNG_corr_profile_z0}), which suggests the same physical origin.

For the most massive halo bin ($M_{200} > 10^{14} \,\msun$), the coupling between gas and dark matter does not vanish completely (Figure~\ref{fig:TNG_corr_profile_z0}, left panel), suggesting that the gravitational coupling is still a major player. Similarly, the anti-correlation between gas and stars does not completely vanish (Figure~\ref{fig:TNG_fb_profile_z0}), which indicates that the feedback cannot eliminate the gravitational coupling between gas and stars  for the most massive systems.
However, recent empirical results of positive correlation between the hot and cold baryons challenged this model \citep[][]{Puddu:2022}.  Since this anti-correlation has been observed across several hydrodynamical simulations \citep[][]{Wu:2015,Farahi:2018}, the empirical evidence of positive correlation can have profound implications for constraining feedback physics implemented in modern hydrodynamical cosmological simulations.

Employing a zeroth-order approximation, the correlation statistics of gas and stellar fraction can be estimated from the density profile statistics (see Appendix~\ref{app:ga_stellar_fraction_statistics} for a derivation). At a fixed normalized radius and mass, the covariance between gas and stellar fraction can be approximated with the scatter in dark matter (denoted with $m$), stellar (denoted with $\star$), and gas (denoted with $g$) density profiles and the correlation across them,
\begin{equation}
 {\rm cov}(f_{g}, f_{\star}) = \sigma_m (\sigma_m - \rgm  \sigma_g) + \sigma_s ( \rgs \sigma_g -  \rsm \sigma_m).
\end{equation}
In the outskirt of halos, the second term in this expression is dominant. Since $\rsm \sigma_m > \rgs \sigma_g$, we expect to get a negative correlation.

\subsection{Limitations and Future Directions}

\begin{figure*}
     \centering
     \includegraphics[width=0.98\textwidth]{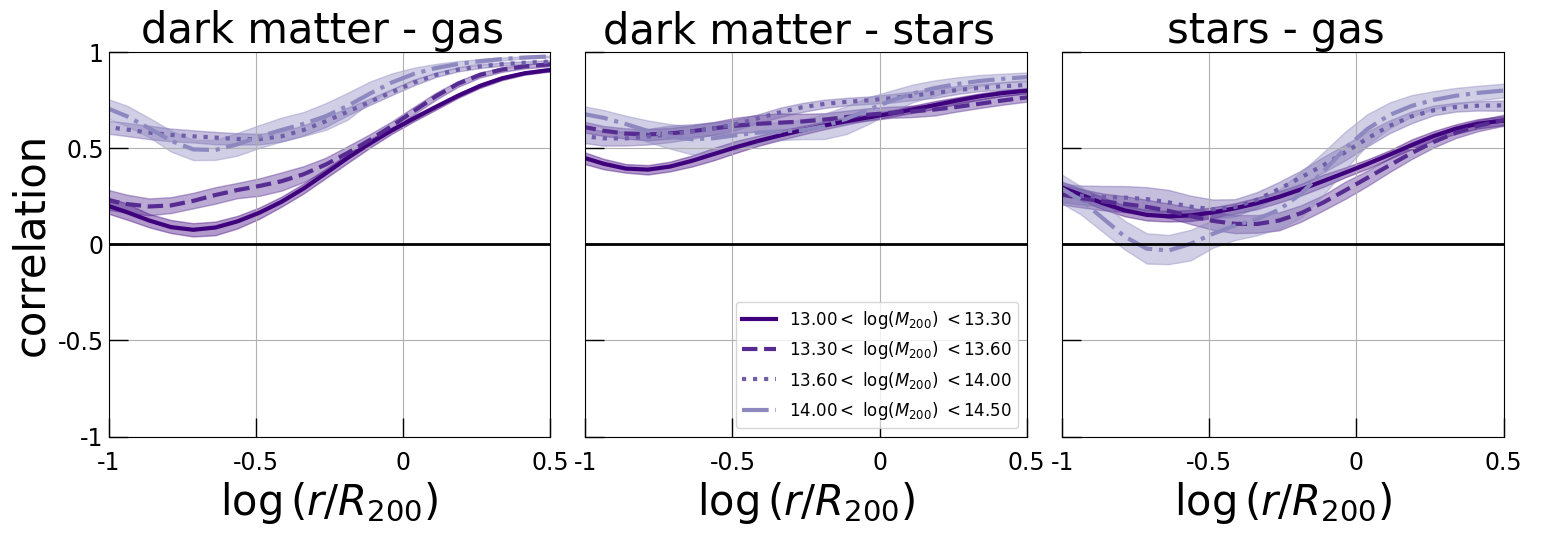} 
\caption{The correlation between the three principle phases of matter (dark matter, gas, and stars) at fixed distance from the center of the halo.}
     \label{fig:TNG_corr_profile_z0}
\end{figure*}

\begin{figure}
     \centering
     \includegraphics[width=0.45\textwidth]{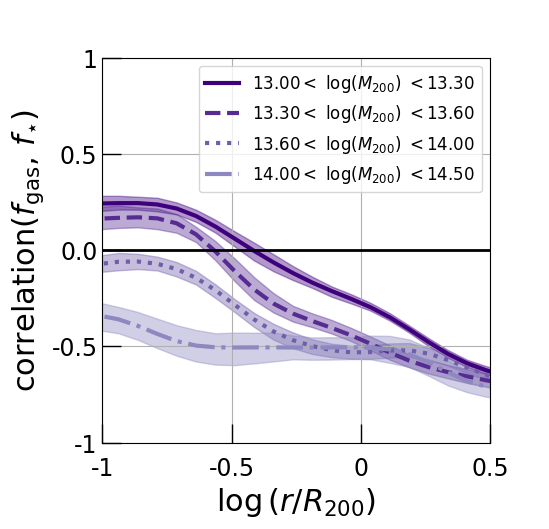} 
\caption{The correlation between differential gas and stellar mass fraction as a function of radial distance from the center of halos. Different lines correspond to halos in different mass bins.}
     \label{fig:TNG_fb_profile_z0}
\end{figure}

The sensitivity of the above results to TNG resolution and box-size are tested in Appendix \ref{app:resolution}, where we demonstrate the level of numerical convergence among TNG simulations. Qualitatively, the results and trends remain intact when switching between different resolutions or box sizes. Quantitatively, the variations in the correlation coefficients of different runs is within a few percent range. In Appendix~\ref{app:TNG_corr_table}, we report the numeric value of correlation coefficients used to make plots in Figure~\ref{fig:TNG_corr_profile_z0} and Figure~\ref{fig:TNG_fb_profile_z0}. It is important to note that with TNG simulations we cannot make a causal connection between feedback physics and the change in the coupling strength of various properties and profiles. Future theoretical studies should establish the potential link between feedback processes and this decoupling scale.

The thermodynamic profiles of galaxy groups and clusters are sensitive to feedback physics, hence vary significantly across modern hydrodynamical simulations \citep[][for a recent review]{Oppenheimer:2021}. For galaxy- and group-size halos, much of this variation originates from the variation in the stellar and AGN feedback \citep[e.g.,][]{McCarthy:2010,Truong:2016,Davies:2019,Terrazas:2020}. How does a more gentle or bursty feedback affect the correlations between gas, stellar, and dark matter density profiles? Can these correlation be used to constrain models of stellar and AGN feedback? How well can we measure and hence constrain feedback model using the correlations of profiles from upcoming multi-wavelength astronomical surveys? The answer to these questions might shed light on the still poorly understood feedback physics and help advance the use of ongoing and upcoming surveys to advance cosmology and galaxy formation.

\section{Discussion} \label{sec:discussion}

Upcoming multi-wavelength cosmological surveys -- such as Rubin, Roman \citep{Rubin-Roman}, eROSITA \citep{Bulbul2021}, Simons Observatory \citep{Ade2019}, CMB-S4 \citep{CMB-S4}, and CMB-HD \citep{CMB-HD} -- will discover a large sample of groups and clusters, which will significantly improve measurements of their abundance \citep[][and references therein]{Allen:2011,Pratt:2019} and auto- and cross-correlation \citep[{\sl e.g.},][]{Osato:2018,Osato:2020,Shirasaki:2020} of dark matter halos over cosmic time. 
However, in order to harness the statistical power of these surveys, we must understand and quantify systematic uncertainties in observable-mass relations and selection function. 
The profile covariance is one of these modeling systematics that is currently overlooked in the literature.
In the following, we discuss the applications of this work to cluster abundance and cross-correlation analysis in the era of multi-wavelength cosmological surveys.

\subsection{Cluster Mass Calibration} \label{sec:cluster_cosmo}

The halo mass-observable relation is one of the key ingredients of cluster abundance analyses. 
Over the past three decades, a number of techniques have been proposed to infer the total mass of the host halos \citep[see][for a recent review]{Pratt:2019}. The weak-lensing method has emerged as one of the promising techniques for the cluster mass calibration \citep[][]{McClintock2019,Kiiveri2021,Wu2021}. 
Past studies have illustrated the sensitivity of inferred halo masses to non-zero scatter \citep[][]{Rozo2014,DES_cluster_cosmo:2020,Grandis:2021}.
In the following, we argue that modeling scatter alone is insufficient for accurate calibration of mass-observable relations for cosmological inference, because halo property profile covariance plays a crucial role in estimating the expected halo mass given an observable.

Clusters are not selected by their host halo mass. Instead, they are selected by their stellar or gas observables. As first pointed out by \citet{Nord:2008} and later shown by others \citep[{\sl e.g.},][]{Farahi2018,2022:Zhang}, a non-zero covariance between observables can induce a non-trivial biases in mass-observable relation if not accounted for properly. As we will show, this bias is not a constant factor, rather it is a  function of mass and radial distance. In the following, we show that this bias can be modelled with an additive term. In the following, we model the cumulative mass profile enclosed within a radial distance from the halo center. We use $M_{<r}$ to denote the cumulative mass profile.

Suppose an optical-survey in which clusters are selected based on their optical-richness, where optical-richness, denoted with $\Ngal$, is defined as the number of satellite galaxies with $M_{\star} > 10^{10}\,\msun$, corresponding to the stellar mass limit of the recent optical-cluster finding algorithms \citep[e.g.,][]{Palmese:2020}.
Since the selection property is $\Ngal$, we want a model that computes expected $M_{<r}$ as a function of redshift, radii, and $\Ngal$, $\langle \ln M_{<r} \mid \Ngal, r \rangle$.
Suppose the conditional probability $p(M_{<r}, \Ngal \mid M)$ is a multivariate log-normal distribution \citep{Huang:2021}; and the $M$--$\Ngal$ scaling relation is given by 
\begin{align} \label{eq:M-Ngal}
  \langle \ln(M) \mid \Ngal, z \rangle = &  \ln(M_0) + \alpha_{\Ngal} \ln\left(\frac{\Ngal}{40}\right) 
  \nonumber\\
  & + \beta_{\Ngal} \ln\left(\frac{1 + z}{1.35}\right), 
\end{align}
where $\ln(M_0)$ is the normalization with the value of $M_0 = 3.081\times 10^{14}\,\,[\msun]$ and $\alpha_{\Ngal}=1.356$ and $\beta_{\Ngal}=-0.30$ are the slopes of the scaling relation, respectively \citep{McClintock2019}. To compute the expected value, we follow the population model of \citet{Evrard:2014}, which approximates the natural log of the halo mass function ${\rm d}n(\mu,z)/{\rm d}\mu$ with a polynomial expression. At a fixed redshift, we employ a first order approximation  
\begin{equation}
{\rm hmf}(\mu) = \frac{{\rm d}n(\mu)}{{\rm d}\mu} \approx A_0 \exp\left[- \gamma (\mu-\mu_0)  \right],    
\end{equation}
where $\mu_0$ is the pivot mass, $A_0$ is the normalization, and $\gamma$ the coefficient of the first order expansion both evaluated at the pivot mass. $\gamma$ can be approximated with
\begin{align} 
    \gamma  &= \left. \frac{\partial \ln {\rm hmf}}{\partial \mu } \right|_{\mu=\mu_0} \\
               &\approx \frac{\ln {\rm hmf}(\mu_0-\delta) - {\ln \rm hmf}(\mu_0+\delta)}{2 \delta}, \nonumber
\end{align}
where $\delta$ is a small number.
The shape of halo mass function encoded in $\gamma$ plays a key role in evaluating the conditional expected value. 
By marginalising over halo mass, the expected cumulative mass profile at fixed optical-richness and redshift is given by 
\begin{align} \label{eq:final_model}
    \langle \ln M_{<r} \mid & \Ngal, z \rangle = \langle \ln M_{<r} \mid M(\Ngal), z \rangle \\
    &   + \gamma \, \alpha_{\Ngal} \times {\rm cov}(\ln M_{<r}  , \ln\Ngal \mid M(\Ngal), z), \nonumber
\end{align}
where $M(\Ngal) \equiv \exp(\langle \ln(M) \mid \Ngal, z \rangle)$ is the mass evaluated at $\Ngal$ by employing the $M$--$\Ngal$ relation in Equation~\eqref{eq:M-Ngal}.  
The first term on the right-hand-side of Equation~\eqref{eq:final_model} is the expected ln-$M_{<r}$ at fixed redshift and halo mass evaluated at $\Ngal$. The second term on the right-hand-side is the covariance between ln-$M_{<r}$ and ln-$\Ngal$ at fixed halo mass and redshift. $\gamma$ and $\alpha_{\Ngal}$ are the logarithmic slopes of the halo mass function at the pivot mass and the slope of $M$--$\Ngal$ relation in Equation~\eqref{eq:M-Ngal}, respectively.

\begin{figure}
     \centering
     \includegraphics[width=0.45\textwidth]{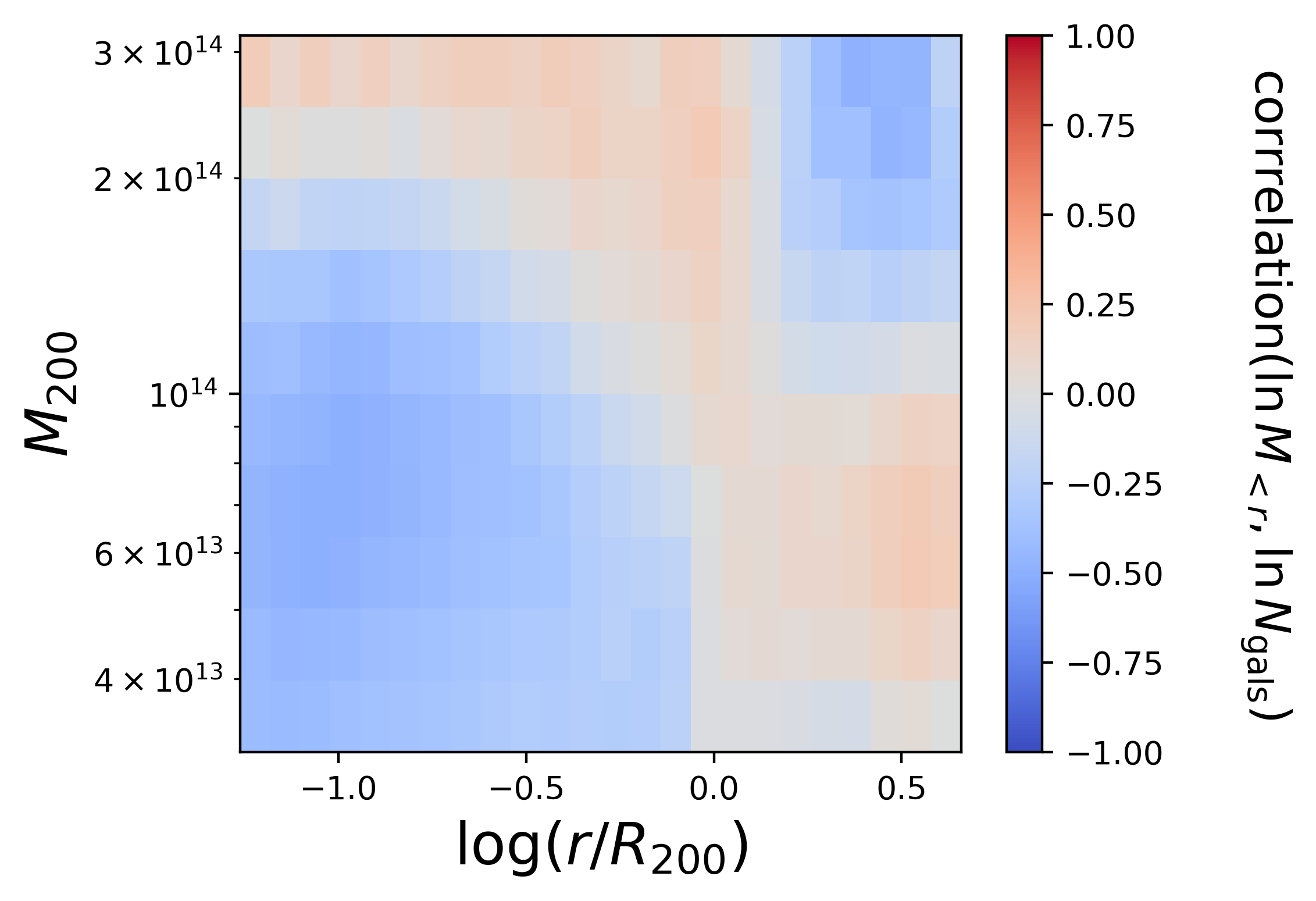}
     \caption{The correlation matrix between $\ln M_{<r}$ and $\ln\Ngal$ as a function of normalized radii $x=r/R_{200}$ and halo mass for halos at redshift $z=0.24$.}
     \label{fig:cov_Ngal_rho}
\end{figure}

Both terms on the right-hand-side of Equation~\eqref{eq:final_model} are functions of radial distance from the halo center, where the radial dependence arises from the expected enclosed mass in the first term and the covariance matrix in the second term.
Figure~\ref{fig:cov_Ngal_rho} demonstrates the radial-dependence of the correlation between the enclosed mass and $\Ngal$ for massive halos at redshift $z=0.24$, a typical median redshift of the modern, low-redshift optical cluster surveys \citep{Costanzi:2019}. Note that, since the bias is proportional to the correlation coefficient, the change in sign of the correlation coefficient determines the sign of bias. We find that the correlation coefficient of group size halos is negative for radial scales less than $R_{200}$, while the same quantity is positive for massive halos. This indicates that not accounting for this systematic can lead to over-estimation of mass in some scales and under-estimation in other scales. 

\begin{figure*}
     \centering
     \includegraphics[width=0.98\textwidth]{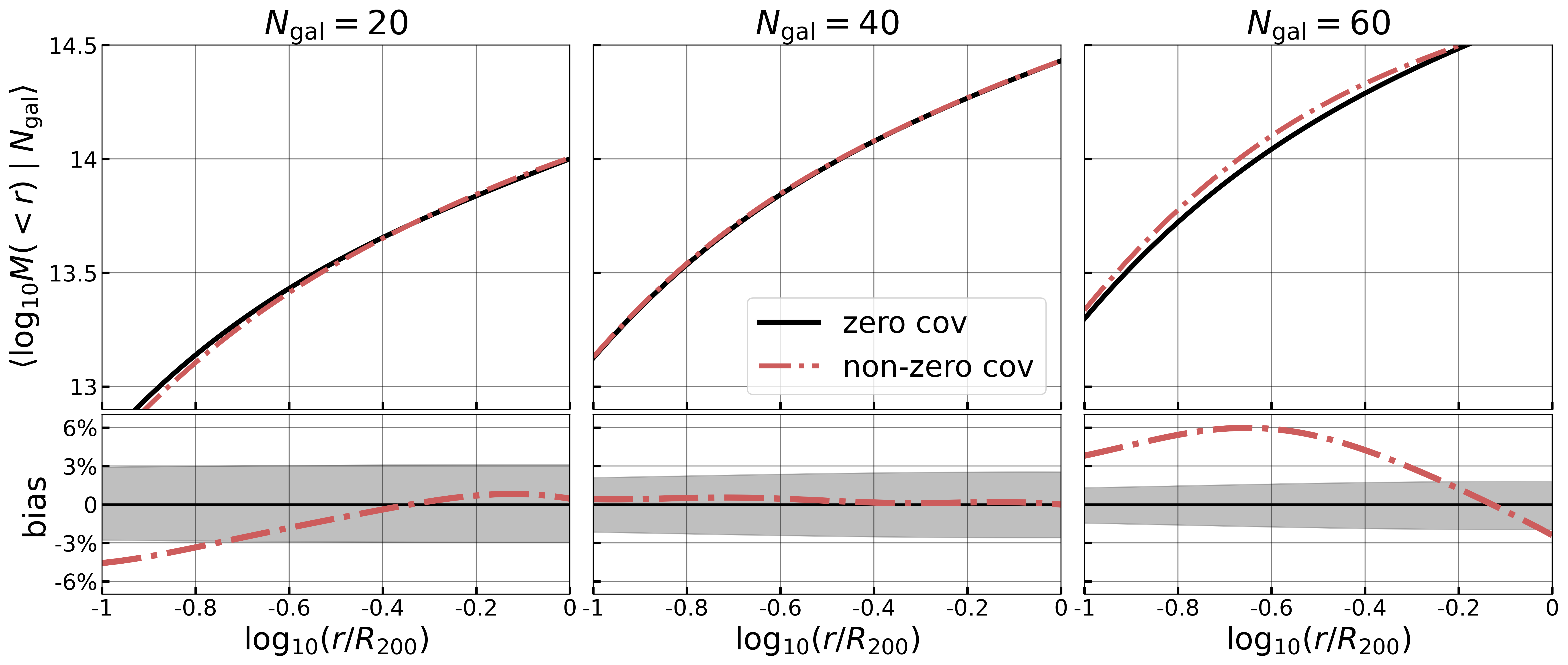} 
\caption{The property covariance bias in cumulative mass profile at fixed optical richness is larger than the reported uncertainties in $N_{\rm gal}$--$M$ relation. This figures shows the expected cumulative mass profile at $N_{\rm gal}=20$, $40$, and $60$ with/without accounting for the property covariance. The shaded region is $68\%$ uncertainty due to the uncertainty in $N_{\rm gal}$--$M$ relation.      }
     \label{fig:expected_bias}
\end{figure*}

Figure~\ref{fig:expected_bias} shows how the correlated scatter introduces bias in the cumulative mass profile through the covariance matrix. The top panels show the expected cumulative mass profile with and without the covariance properly taken into account for three optical-richness bins. The bottom panels show the fractional bias induced by the radial-dependent covariance in the TNG simulation. The shaded region shows the $68\%$ uncertainty region due to the uncertainty in the $\Ngal$--$M$ scaling relation reported in Table~6 of \citet{McClintock2019}. We notice that the bias due to the property covariance can be as large as $5\%$, indicating that this systematic uncertainty is as large as other systematics noted in \citet{McClintock2019} and cannot be neglected in all future mass calibration analyses. Since this bias is both mass- and radial-dependent in weak-lensing analyses, one cannot employ a multiplicative or additive correction in a post-analysis to account for this bias.  

This work studies the 3D mass profile of dark matter halos selected on the true number of galaxies. In observational studies, neither of these quantities is observable. Real observables are the projected quantities in RA, DEC, and redshift space; thus, there will be additional contributions to the covariance matrix due to the projection effect that is not modeled in this work \citep[][]{Costanzi:2019,2022:Zhang}. Additionally, the stellar property statistics of dark matter halos, including $\Ngal$, deviates from a log-normal distribution \citep[][]{Anbajagane:2020}. These systematic uncertainties must be quantified in a future study.

\subsection{Forward-modeling Multi-wavelength Cross-Correlation Signal} \label{sec:cluster_cosmo}

In the era of multi-wavelength cosmological surveys, cross-correlation is one of the most powerful techniques for constraining cosmology and astrophysics. Cross-power spectra of any two fields, such as the tSZ effect, X-ray surface brightness, and lensing convergence maps, promises to provide novel and complementary cosmological and astrophysical constraints \citep[{\sl e.g.},][]{Shirasaki:2020, Pandey2020tSZForecast}. These signals are commonly modelled using the halo-model approach, which takes as an input the mean halo profiles of dark matter, gas, and stellar properties. However, modern halo-model approaches often assume baryons (gas and galaxies) trace the underlying dark matter profiles and also do not account for the non-trivial correlation among these components.

In this work, we found that there are significant scatter and de-correlation among the dark matter, gas, and stellar profiles, especially in the interior of low-mass halos that are impacted by feedback effects. We showed that the scatter and the degree of (de-)correlation is strongly radial and mass dependent. These effects will introduce  biases in modeling and interpretation of the cross-correlation studies, if these effects are not properly taken into account. In our future work, we plan to assess the magnitude of this bias by incorporating the scatter and correlations in the Baryon Pasting model \citep[{\sl e.g.},][]{Osato2022}. 

\section{Conclusions} \label{sec:conclusion}

In this work, we study the radial-dependence of coupling between dark matter, gas, and stellar log-density profiles as a function of halo mass and cluster-centric distances. Our key findings are summarized in the following points.

\begin{itemize}
\item The correlation between scatter of stellar, hot gas, and dark matter density profiles are mass- and radial-dependent.
\item At scales larger than $R_{200}$, there is a strong positive correlation between dark matter, gas, and stellar density profiles, implying that all components are tightly coupled.
\item At scales less than $0.3\,R_{200}$, the dark matter density is still coupled to stars, while the correlation between dark matter density and gas density fades. 
\item The gravitational coupling of baryons and dark matter at large radii and decoupling at small radii can be also probed with the anti-correlation and correlation, respectively, between $f_{\rm gas}$ and $f_{\star}$.  
\item The covariance between halo properties induces radial-dependent additive bias in the expected cumulative mass profile of optically-selected clusters.  
\end{itemize}

In this work, we studied the radial-dependence of coupling strength between the halo density profiles. While we speculate that energy injection by SNe and AGN would be the primary mechanism behind the dark matter--gas gravitational decoupling; however, more in-depth analysis is required to establish such a causal relation. Another interesting direction for a future work would be a thorough stress-testing of the correlations to changes in sub-grid physics and cosmology.  Simulations such as the CAMELS project, a suite of simulations that systematically vary the sub-grid physics and the cosmological parameters, will enable such studies \citep{Villaescusa-Navarro2021,CAMELS:2022data_release}.

Upcoming multi-wavelength cluster surveys -- such as Rubin, Roman, eROSITA, Simons Observatory, CMB-S4, and CMB-HD -- are designed to significantly improve cosmological and astrophysical constraints through the combination of measurements of individual halos as well as cross-correlation studies. 
Scale-dependent scatter and correlation in halo profiles induce non-trivial biases in modeling and interpretation of the cluster abundance cosmology and cross-correlation studies.
Modeling the halo profile covariance is one of the outstanding challenges facing multi-wavelength cross-correlation studies.
Future works should assess the form and magnitude of these biases by incorporating the scatter and correlations into the predictive models.


\acknowledgments
\noindent {\bf Acknowledgments.} 
The authors would like to thank August~Evrard for useful discussion on the model described in Appendix~\ref{app:ga_stellar_fraction_statistics}. We thank the anonymous referee for constructive feedback that improved the presentation of this paper.
The authors would like to thank David Barnes for his contribution to this work before a career move.  This work is supported by the University of Texas at Austin and Yale University. DA is supported by the National Science Foundation Graduate Research Fellowship under Grant No. DGE 1746045.

\bibliography{mybib}

\appendix

 \section{Redfshift of the Property Profile Correlations} \label{app:z_evo_prop_corr}

\begin{figure*}
     \centering
     \includegraphics[width=0.99\textwidth]{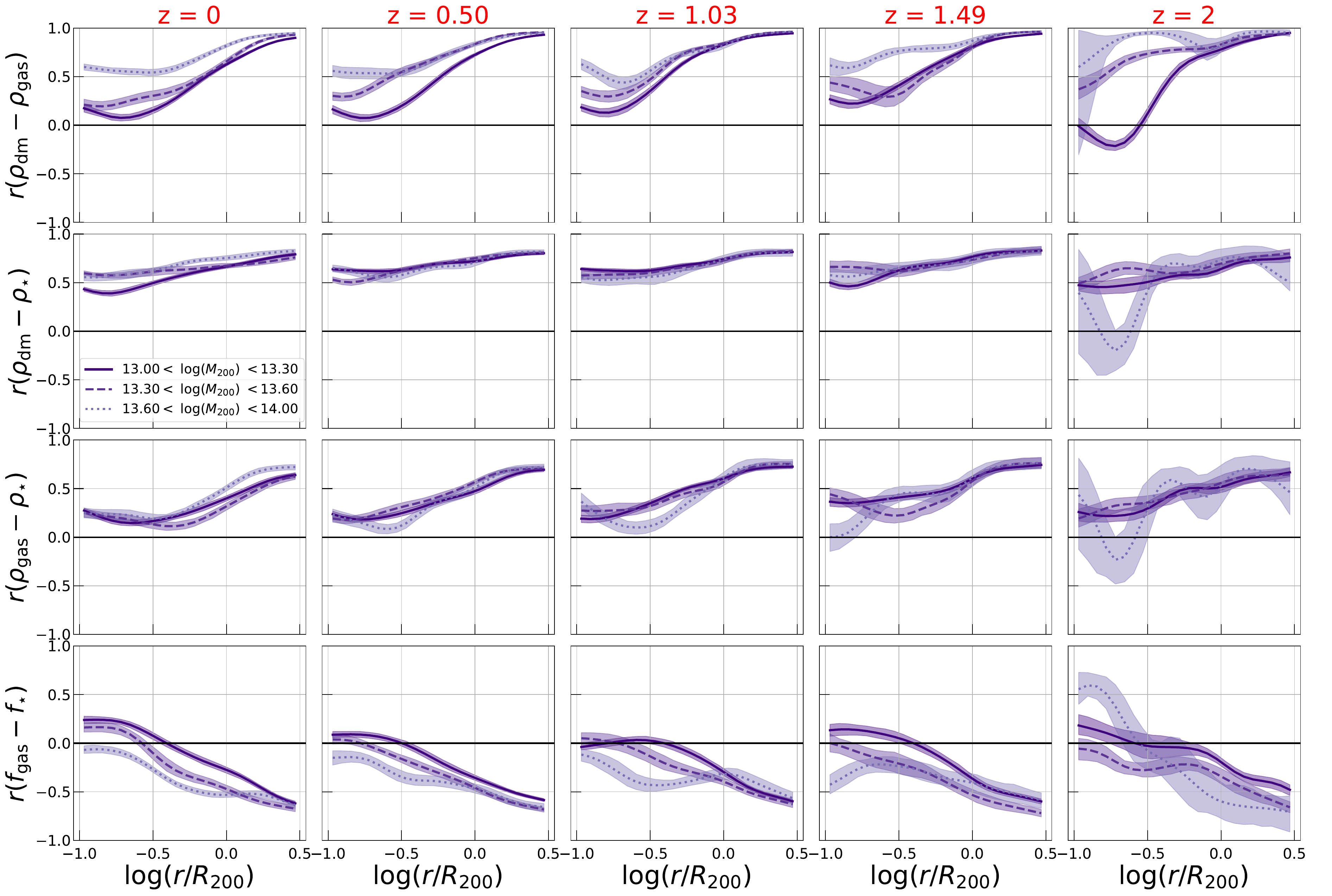} 
\caption{The top three rows are the correlation between the dark matter, gas, and stars differential densities at fixed distance from the center of halo in 3 mass bins. The bottom row is the correlation between differential gas and stellar mass fraction. The columns from left to right shows these correlations at redshifts 0, 0.5, 1.03, 1.49, and 2, respectively. }
     \label{fig:redshift_evolution_TNG}
\end{figure*}

 We present the redshift evolution of density profiles and baryon fraction profiles correlation using the outputs of TNG-100 simulations in Figure \ref{fig:redshift_evolution_TNG}. We notice the correlation profiles have a weak dependency to redshift. Since the uncertainties also increase, we avoid drawing a conclusion. The general trends remain intact at all redshifts for all the correlations up to redshift 2.  A noticeable feature is that the correlation between $\rho_{\rm dm}$ and $\rho_{\rm gas}$ starts to show a dip at scales around $0.3 R_{200}$ about redshift 1 for the lowest mass halo.

\section{Simulation resolution} \label{app:resolution}

\begin{figure}
     \centering
     \includegraphics[width=0.80\textwidth]{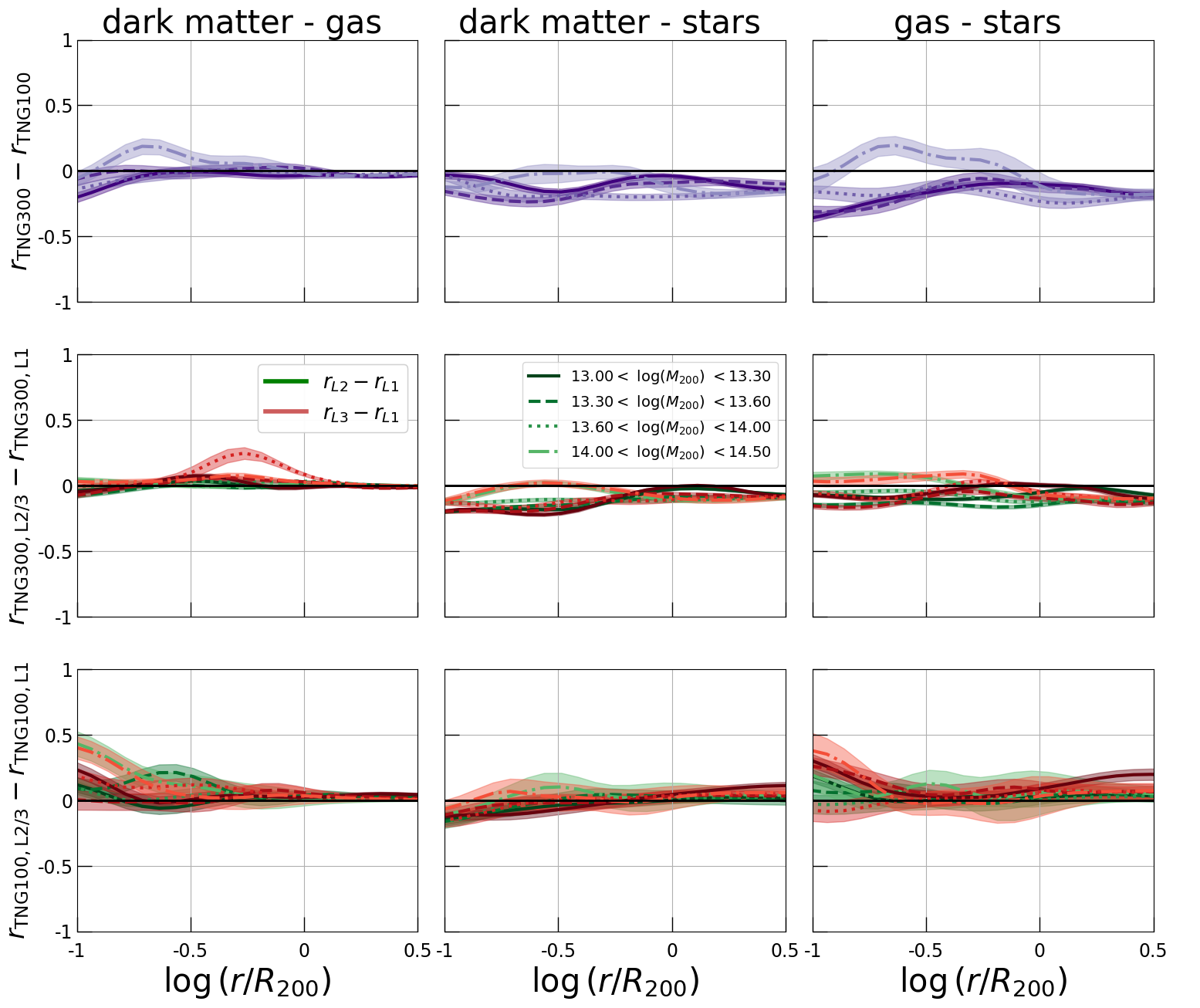} \\ 
\caption{ Impact of resolution and simulation box on the correlation between dark matter, gas, and stars density profiles. The panels shows the difference between correlations measured from simulations of different box size (purple lines) or resolution (red and green lines). In most cases the correlation statistics are consistent with the statistical uncertainty of the sample, except for the gas and dark matter correlation of the most massive bin of halos in TNG300 vs TNG100 simulations. The label L1 (L3) corresponds to the highest (lowest) resolution TNG simulation. }
     \label{fig:resolution_dependence}
\end{figure}

Figure \ref{fig:resolution_dependence} illustrates the sensitivity of the density profiles correlation signals for three resolutions (top to bottom) of TNG100 with $110.7^3\, {\rm Mpc}^3$ (left panels) and TNG300 with $302.6^3\, {\rm Mpc}^3$ (right panels) box sizes. In our main analysis, we employ TNG100 since its fiducial run is $8$ times better in resolution than TNG300 fiducial run. The trends and the gas--dark matter decoupling scales are independent of the simulations resolution and box size, and the results are in good agreement. The actual value of correlations has $< 10\%$ sensitivity to the box size, specially for the largest halo mass bin. This might be linked to the sample selection caused by the cosmic variance. Halos in the most massive bin of the TNG100 box are not fully representative of the population of similar-mass halos in the TNG300 box.

\section{Statistics of Gas and Stellar Fractions} \label{app:ga_stellar_fraction_statistics}

Consider the density profile components for dark matter ($\rhom$), gas ($\rhog$) and stars ($\rhos$) as a function of normalized radius within an ensemble of halos. We employ a multivariate log-normal model for all halo properties. Let $\langle \cdots \rangle$ denote the expected value conditioned on $M_{\rm halo}$ at some radius and epoch. Then, at a fixed radius and epoch, each halo has a deviation around the mean log-scaling relations in gas, stars, and dark matter: 
\begin{eqnarray}
    \dg &=& \ln \rhog - \langle \ln \rhog \rangle, \\ 
    \ds &=& \ln \rhos - \langle \ln \rhos\rangle, \\
    \dm &=& \ln \rhom - \langle \ln \rhom \rangle . 
\end{eqnarray}
At each radius, if dark matter is dominant, then the gas and stellar mass fraction can be approximated as: 
\begin{equation}
    \ln \fg = \ln \rhog - \ln ( {\rhom + \rhog + \rhos} ) \simeq \ln \rhog - \ln \rhom,
    \label{eq:approx}
\end{equation}
where the second term with $\ln(\rhom + \rhog + \rhos)$, can be approximated as $\ln ( {\rhom} )$ for $\rhom \gg (\rhog + \rhos)$. 
To the extent that the approximation in Equation~\ref{eq:approx} holds, then it also holds for the ensemble mean. Each halo then has deviation of gas and stellar mass fractions of 
\begin{eqnarray}
    \dfg &=& \dg - \dm, \\
    \dfs &=& \ds - \dm
\end{eqnarray}
from the mean relation.

The variances in gas and stellar mass fractions, denoted as $\sigma^2_{f,g}$ and $\sigma^2_{f,\star}$ respectively, are given by 
\begin{eqnarray}
    \sigma^2_{f,g} = \langle (\dg - \dm)^2 \rangle  = \sigma^2_m + \sigma^2_g -2 \rgm \sigma_g \sigma_m , \\ 
    \sigma^2_{f,s} = \langle (\ds - \dm)^2 \rangle  = \sigma^2_m + \sigma^2_s -2 \rsm \sigma_s \sigma_m,
\end{eqnarray}
where $\sigma^2_m  \equiv \langle \dm^2 \rangle $, $\sigma^2_{\star}  \equiv \langle \ds^2 \rangle$, $\sigma^2_{g}  \equiv \langle \dg^2 \rangle$, and the correlation coefficients are defined in the usual way and are shown in Figure~\ref{fig:TNG_corr_profile_z0}. The cross-term can then be computed
\begin{equation}
 \langle \dfg \, \dfs \rangle = \sigma^2_m + \rgs \sigma_g \sigma_s - \rgm  \sigma_g \sigma_m - \rsm  \sigma_s \sigma_m,
 \label{eq:crosscoeff}
\end{equation}
where the compute correlation coefficient is given by
\begin{equation}
 r_{f,g\star} = \frac{ \langle \dfg \, \dfs \rangle }{\sqrt{\langle \dfg \, \dfg \rangle \langle \dfs \, \dfs \rangle}}.
\end{equation}

Figure~\ref{fig:TNG_corr_profile_z0} shows that all correlations coefficients ($\rgm, \rsm, \rgm$) are positive at all radii, demonstrating that one can still get a negative covariance in the component mass fractions as long as the last two terms dominate over the first two in Equation~\ref{eq:crosscoeff}. Finally, we test this model using the highest mass halo bin. Table~\ref{tab:my_label} reports the estimated parameters at two different radii, which shows the anti-correlation both in the interior and exterior of the dark matter halo.

\begin{table}[t!]
    \centering
    \begin{tabular}{|c||c|c|c|c|c|c||c|c|}
    \hline
    $r/R$ & $\sigma_m$ & $\sigma_{\star}$ & $\sigma_g$ & $\rgs$ & $ \rgm $ & $\rsm$  & $r_{f,g\star}$ & Figure 4 \cr
    \hline 
    $0.10$ & $0.12$ & $0.26$ & $0.16$ & $0.29$ & $0.71$ & $0.67$ & $-0.36$ & $-0.34$  \cr
    $3.07$ & $0.29$ & $0.59$ & $0.24$ & $0.79$ & $0.97$ & $0.87$ & $-0.68$ & $-0.71$ \cr
    \hline
    \end{tabular}
    \caption{Estimated parameters (first six) and derived (last four) quantities.  The one before the last column should be compared with the values reported in Figure~\ref{fig:TNG_fb_profile_z0}. } 
    \label{tab:my_label}
\end{table}

\section{Correlation Data} \label{app:TNG_corr_table}

In Table~\ref{tab:corr_table}, we report the correlation coefficient between halo property profiles for halos in four mass bins at redshift $0$. These data are used to make plots in Figure~\ref{fig:TNG_corr_profile_z0} and Figure~\ref{fig:TNG_fb_profile_z0}. The top and bottom tables present the results for TNG100 and TNG300, respectively.   These correlation values with three significant digits will be available in the electronic version.

\newpage 
\pagebreak[4]
\global\pdfpageattr\expandafter{\the\pdfpageattr/Rotate 90}

\begin{sidewaystable}[]
    \caption{The correlation coefficient between scatter of halo profiles for four halo mass bins. Top (bottom) table reports the results of TNG100 (TNG300) simulations at redshift zero.}
\label{tab:corr_table}
{\tiny
\begin{tabular}{|l|c|c|c|c|c|c|c|c|c|c|c|c|c|}
\hline 
\multicolumn{1}{|c|}{Profiles} & $\log(M\, [M_{\odot}])$ & \multicolumn{12}{c|}{$x = r/R_{200c}$} \\ \hline
 &  & $ 0.07$ & $ 0.10$ & $ 0.14$ & $ 0.19$ & $ 0.27$ & $ 0.39$ & $ 0.55$ & $ 0.77$ & $ 1.09$ & $ 1.54$ & $ 2.18$ & $ 3.07$ \\
\hline
 $\rho_{\rm dm} - \rho_{\rm gas}$ & $[13.00, 13.30]$ & $ 0.23_{\pm0.04}$ & $ 0.20_{\pm0.04}$ & $ 0.12_{\pm0.03}$ & $ 0.07_{\pm0.03}$ & $ 0.11_{\pm0.03}$ & $ 0.22_{\pm0.03}$ & $ 0.37_{\pm0.02}$ & $ 0.53_{\pm0.02}$ & $ 0.66_{\pm0.02}$ & $ 0.78_{\pm0.01}$ & $ 0.86_{\pm0.01}$ & $ 0.90_{\pm0.01}$ \\
  & $[13.30, 13.60]$ & $ 0.27_{\pm0.08}$ & $ 0.22_{\pm0.06}$ & $ 0.19_{\pm0.05}$ & $ 0.22_{\pm0.05}$ & $ 0.28_{\pm0.05}$ & $ 0.32_{\pm0.05}$ & $ 0.41_{\pm0.05}$ & $ 0.54_{\pm0.03}$ & $ 0.69_{\pm0.02}$ & $ 0.83_{\pm0.02}$ & $ 0.91_{\pm0.01}$ & $ 0.93_{\pm0.01}$ \\
  & $[13.60, 14.00]$ & $ 0.62_{\pm0.05}$ & $ 0.60_{\pm0.04}$ & $ 0.57_{\pm0.04}$ & $ 0.55_{\pm0.04}$ & $ 0.54_{\pm0.03}$ & $ 0.56_{\pm0.03}$ & $ 0.64_{\pm0.02}$ & $ 0.74_{\pm0.02}$ & $ 0.84_{\pm0.01}$ & $ 0.91_{\pm0.01}$ & $ 0.94_{\pm0.01}$ & $ 0.95_{\pm0.01}$ \\
  & $[14.00, 14.50]$ & $ 0.76_{\pm0.05}$ & $ 0.71_{\pm0.06}$ & $ 0.60_{\pm0.05}$ & $ 0.50_{\pm0.06}$ & $ 0.52_{\pm0.06}$ & $ 0.60_{\pm0.06}$ & $ 0.68_{\pm0.07}$ & $ 0.79_{\pm0.06}$ & $ 0.89_{\pm0.03}$ & $ 0.93_{\pm0.02}$ & $ 0.96_{\pm0.01}$ & $ 0.97_{\pm0.01}$ \\
\hline
 $\rho_{\rm dm} - \rho_{\star}$ & $[13.00, 13.30]$ & $ 0.52_{\pm0.03}$ & $ 0.45_{\pm0.03}$ & $ 0.39_{\pm0.03}$ & $ 0.40_{\pm0.03}$ & $ 0.47_{\pm0.03}$ & $ 0.53_{\pm0.02}$ & $ 0.59_{\pm0.02}$ & $ 0.63_{\pm0.02}$ & $ 0.67_{\pm0.02}$ & $ 0.72_{\pm0.02}$ & $ 0.77_{\pm0.01}$ & $ 0.80_{\pm0.01}$ \\
  & $[13.30, 13.60]$ & $ 0.65_{\pm0.03}$ & $ 0.60_{\pm0.03}$ & $ 0.57_{\pm0.03}$ & $ 0.57_{\pm0.04}$ & $ 0.60_{\pm0.04}$ & $ 0.62_{\pm0.03}$ & $ 0.64_{\pm0.03}$ & $ 0.66_{\pm0.03}$ & $ 0.68_{\pm0.03}$ & $ 0.69_{\pm0.03}$ & $ 0.73_{\pm0.02}$ & $ 0.76_{\pm0.02}$ \\
  & $[13.60, 14.00]$ & $ 0.61_{\pm0.04}$ & $ 0.56_{\pm0.04}$ & $ 0.55_{\pm0.04}$ & $ 0.58_{\pm0.04}$ & $ 0.60_{\pm0.04}$ & $ 0.65_{\pm0.03}$ & $ 0.71_{\pm0.02}$ & $ 0.74_{\pm0.02}$ & $ 0.76_{\pm0.02}$ & $ 0.79_{\pm0.02}$ & $ 0.81_{\pm0.02}$ & $ 0.83_{\pm0.02}$ \\
  & $[14.00, 14.50]$ & $ 0.68_{\pm0.06}$ & $ 0.67_{\pm0.05}$ & $ 0.62_{\pm0.05}$ & $ 0.55_{\pm0.07}$ & $ 0.55_{\pm0.07}$ & $ 0.57_{\pm0.06}$ & $ 0.59_{\pm0.07}$ & $ 0.65_{\pm0.07}$ & $ 0.75_{\pm0.06}$ & $ 0.81_{\pm0.04}$ & $ 0.85_{\pm0.03}$ & $ 0.87_{\pm0.03}$ \\
\hline
 $\rho_{\rm gas} - \rho_{\star} $ &  $[13.00, 13.30]$ & $ 0.36_{\pm0.04}$ & $ 0.30_{\pm0.03}$ & $ 0.22_{\pm0.03}$ & $ 0.16_{\pm0.03}$ & $ 0.15_{\pm0.03}$ & $ 0.19_{\pm0.03}$ & $ 0.25_{\pm0.03}$ & $ 0.33_{\pm0.03}$ & $ 0.42_{\pm0.03}$ & $ 0.52_{\pm0.02}$ & $ 0.60_{\pm0.02}$ & $ 0.64_{\pm0.02}$ \\
  & $[13.30, 13.60]$ & $ 0.31_{\pm0.05}$ & $ 0.27_{\pm0.04}$ & $ 0.23_{\pm0.04}$ & $ 0.20_{\pm0.04}$ & $ 0.15_{\pm0.05}$ & $ 0.11_{\pm0.05}$ & $ 0.13_{\pm0.04}$ & $ 0.22_{\pm0.04}$ & $ 0.35_{\pm0.04}$ & $ 0.47_{\pm0.04}$ & $ 0.58_{\pm0.04}$ & $ 0.64_{\pm0.03}$ \\
  & $[13.60, 14.00]$ & $ 0.30_{\pm0.05}$ & $ 0.26_{\pm0.05}$ & $ 0.24_{\pm0.05}$ & $ 0.23_{\pm0.05}$ & $ 0.20_{\pm0.05}$ & $ 0.21_{\pm0.04}$ & $ 0.29_{\pm0.04}$ & $ 0.41_{\pm0.04}$ & $ 0.55_{\pm0.04}$ & $ 0.66_{\pm0.03}$ & $ 0.71_{\pm0.02}$ & $ 0.72_{\pm0.02}$ \\
  & $[14.00, 14.50]$ & $ 0.34_{\pm0.09}$ & $ 0.29_{\pm0.08}$ & $ 0.13_{\pm0.08}$ & $ -0.02_{\pm0.08}$ & $ 0.01_{\pm0.07}$ & $ 0.11_{\pm0.07}$ & $ 0.18_{\pm0.08}$ & $ 0.36_{\pm0.11}$ & $ 0.57_{\pm0.09}$ & $ 0.70_{\pm0.06}$ & $ 0.76_{\pm0.04}$ & $ 0.79_{\pm0.04}$ \\
\hline
 $f_{\rm gas} - f_{\star}$  & $[13.00, 13.30]$ & $ 0.24_{\pm0.05}$ & $ 0.24_{\pm0.04}$ & $ 0.24_{\pm0.04}$ & $ 0.22_{\pm0.03}$ & $ 0.12_{\pm0.03}$ & $ -0.00_{\pm0.04}$ & $ -0.12_{\pm0.04}$ & $ -0.21_{\pm0.03}$ & $ -0.30_{\pm0.02}$ & $ -0.41_{\pm0.02}$ & $ -0.54_{\pm0.02}$ & $ -0.63_{\pm0.02}$ \\
  & $[13.30, 13.60]$ & $ 0.16_{\pm0.07}$ & $ 0.16_{\pm0.06}$ & $ 0.17_{\pm0.05}$ & $ 0.14_{\pm0.05}$ & $ -0.01_{\pm0.05}$ & $ -0.20_{\pm0.06}$ & $ -0.33_{\pm0.04}$ & $ -0.40_{\pm0.04}$ & $ -0.49_{\pm0.04}$ & $ -0.57_{\pm0.04}$ & $ -0.63_{\pm0.03}$ & $ -0.68_{\pm0.03}$ \\
  & $[13.60, 14.00]$ & $ -0.11_{\pm0.05}$ & $ -0.07_{\pm0.05}$ & $ -0.06_{\pm0.04}$ & $ -0.10_{\pm0.04}$ & $ -0.21_{\pm0.04}$ & $ -0.36_{\pm0.04}$ & $ -0.47_{\pm0.04}$ & $ -0.52_{\pm0.04}$ & $ -0.53_{\pm0.03}$ & $ -0.52_{\pm0.03}$ & $ -0.57_{\pm0.03}$ & $ -0.65_{\pm0.03}$ \\
  & $[14.00, 14.50]$ & $ -0.35_{\pm0.09}$ & $ -0.34_{\pm0.08}$ & $ -0.39_{\pm0.07}$ & $ -0.47_{\pm0.07}$ & $ -0.51_{\pm0.08}$ & $ -0.51_{\pm0.08}$ & $ -0.51_{\pm0.06}$ & $ -0.51_{\pm0.06}$ & $ -0.51_{\pm0.06}$ & $ -0.54_{\pm0.07}$ & $ -0.64_{\pm0.07}$ & $ -0.71_{\pm0.05}$ \\
\hline
\hline
 $\rho_{\rm dm} - \rho_{\rm gas}$ & $[13.00, 13.30]$ & $ -0.02_{\pm0.01}$ & $ -0.01_{\pm0.01}$ & $ 0.00_{\pm0.01}$ & $ 0.03_{\pm0.01}$ & $ 0.10_{\pm0.01}$ & $ 0.21_{\pm0.01}$ & $ 0.34_{\pm0.01}$ & $ 0.49_{\pm0.01}$ & $ 0.63_{\pm0.01}$ & $ 0.74_{\pm0.00}$ & $ 0.82_{\pm0.00}$ & $ 0.87_{\pm0.00}$ \\
  & $[13.30, 13.60]$ & $ 0.15_{\pm0.01}$ & $ 0.17_{\pm0.01}$ & $ 0.19_{\pm0.01}$ & $ 0.23_{\pm0.01}$ & $ 0.27_{\pm0.01}$ & $ 0.32_{\pm0.01}$ & $ 0.43_{\pm0.01}$ & $ 0.57_{\pm0.01}$ & $ 0.71_{\pm0.00}$ & $ 0.81_{\pm0.00}$ & $ 0.87_{\pm0.00}$ & $ 0.90_{\pm0.00}$ \\
  & $[13.60, 14.00]$ & $ 0.45_{\pm0.01}$ & $ 0.46_{\pm0.01}$ & $ 0.49_{\pm0.01}$ & $ 0.51_{\pm0.02}$ & $ 0.53_{\pm0.03}$ & $ 0.57_{\pm0.03}$ & $ 0.64_{\pm0.03}$ & $ 0.73_{\pm0.01}$ & $ 0.82_{\pm0.01}$ & $ 0.88_{\pm0.00}$ & $ 0.91_{\pm0.00}$ & $ 0.93_{\pm0.00}$ \\
  & $[14.00, 14.50]$ & $ 0.63_{\pm0.02}$ & $ 0.64_{\pm0.02}$ & $ 0.67_{\pm0.01}$ & $ 0.67_{\pm0.01}$ & $ 0.65_{\pm0.01}$ & $ 0.65_{\pm0.01}$ & $ 0.73_{\pm0.01}$ & $ 0.81_{\pm0.01}$ & $ 0.87_{\pm0.01}$ & $ 0.90_{\pm0.00}$ & $ 0.93_{\pm0.00}$ & $ 0.95_{\pm0.00}$ \\
\hline
 $\rho_{\rm dm} - \rho_{\star}$ & $[13.00, 13.30]$ & $ 0.49_{\pm0.01}$ & $ 0.42_{\pm0.01}$ & $ 0.34_{\pm0.01}$ & $ 0.31_{\pm0.01}$ & $ 0.32_{\pm0.01}$ & $ 0.39_{\pm0.01}$ & $ 0.50_{\pm0.01}$ & $ 0.59_{\pm0.01}$ & $ 0.63_{\pm0.01}$ & $ 0.64_{\pm0.01}$ & $ 0.64_{\pm0.01}$ & $ 0.65_{\pm0.01}$ \\
  & $[13.30, 13.60]$ & $ 0.54_{\pm0.01}$ & $ 0.46_{\pm0.01}$ & $ 0.37_{\pm0.01}$ & $ 0.34_{\pm0.01}$ & $ 0.37_{\pm0.01}$ & $ 0.44_{\pm0.01}$ & $ 0.51_{\pm0.01}$ & $ 0.56_{\pm0.01}$ & $ 0.59_{\pm0.01}$ & $ 0.61_{\pm0.01}$ & $ 0.64_{\pm0.01}$ & $ 0.66_{\pm0.01}$ \\
  & $[13.60, 14.00]$ & $ 0.63_{\pm0.01}$ & $ 0.54_{\pm0.01}$ & $ 0.44_{\pm0.01}$ & $ 0.40_{\pm0.01}$ & $ 0.42_{\pm0.01}$ & $ 0.47_{\pm0.01}$ & $ 0.51_{\pm0.02}$ & $ 0.54_{\pm0.01}$ & $ 0.56_{\pm0.01}$ & $ 0.60_{\pm0.01}$ & $ 0.65_{\pm0.01}$ & $ 0.68_{\pm0.01}$ \\
  & $[14.00, 14.50]$ & $ 0.61_{\pm0.02}$ & $ 0.55_{\pm0.02}$ & $ 0.51_{\pm0.02}$ & $ 0.50_{\pm0.02}$ & $ 0.53_{\pm0.02}$ & $ 0.56_{\pm0.01}$ & $ 0.59_{\pm0.01}$ & $ 0.61_{\pm0.01}$ & $ 0.62_{\pm0.01}$ & $ 0.65_{\pm0.01}$ & $ 0.68_{\pm0.01}$ & $ 0.71_{\pm0.01}$ \\
\hline
 $\rho_{\rm gas} - \rho_{\star} $ & $[13.00, 13.30]$ & $ -0.03_{\pm0.01}$ & $ -0.06_{\pm0.01}$ & $ -0.08_{\pm0.01}$ & $ -0.07_{\pm0.01}$ & $ -0.03_{\pm0.01}$ & $ 0.05_{\pm0.01}$ & $ 0.15_{\pm0.01}$ & $ 0.24_{\pm0.01}$ & $ 0.31_{\pm0.01}$ & $ 0.40_{\pm0.01}$ & $ 0.44_{\pm0.01}$ & $ 0.46_{\pm0.01}$ \\
  & $[13.30, 13.60]$ & $ 0.01_{\pm0.01}$ & $ -0.04_{\pm0.01}$ & $ -0.08_{\pm0.01}$ & $ -0.08_{\pm0.01}$ & $ -0.04_{\pm0.01}$ & $ 0.02_{\pm0.01}$ & $ 0.08_{\pm0.01}$ & $ 0.14_{\pm0.01}$ & $ 0.23_{\pm0.01}$ & $ 0.34_{\pm0.01}$ & $ 0.42_{\pm0.01}$ & $ 0.46_{\pm0.02}$ \\
  & $[13.60, 14.00]$ & $ 0.14_{\pm0.02}$ & $ 0.09_{\pm0.02}$ & $ 0.05_{\pm0.01}$ & $ 0.03_{\pm0.01}$ & $ 0.04_{\pm0.01}$ & $ 0.07_{\pm0.01}$ & $ 0.13_{\pm0.01}$ & $ 0.21_{\pm0.01}$ & $ 0.30_{\pm0.01}$ & $ 0.42_{\pm0.01}$ & $ 0.50_{\pm0.01}$ & $ 0.54_{\pm0.01}$ \\
  & $[14.00, 14.50]$ & $ 0.23_{\pm0.03}$ & $ 0.20_{\pm0.02}$ & $ 0.17_{\pm0.02}$ & $ 0.17_{\pm0.02}$ & $ 0.18_{\pm0.02}$ & $ 0.21_{\pm0.02}$ & $ 0.28_{\pm0.02}$ & $ 0.38_{\pm0.02}$ & $ 0.47_{\pm0.02}$ & $ 0.54_{\pm0.02}$ & $ 0.59_{\pm0.02}$ & $ 0.61_{\pm0.02}$ \\
\hline
 $f_{\rm gas} - f_{\star}$  & $[13.00, 13.30]$ & $ 0.04_{\pm0.02}$ & $ 0.02_{\pm0.01}$ & $ 0.03_{\pm0.01}$ & $ 0.05_{\pm0.01}$ & $ 0.05_{\pm0.01}$ & $ 0.02_{\pm0.01}$ & $ -0.03_{\pm0.01}$ & $ -0.09_{\pm0.01}$ & $ -0.14_{\pm0.01}$ & $ -0.21_{\pm0.01}$ & $ -0.29_{\pm0.01}$ & $ -0.37_{\pm0.01}$ \\
  & $[13.30, 13.60]$ & $ 0.04_{\pm0.02}$ & $ 0.02_{\pm0.02}$ & $ 0.01_{\pm0.01}$ & $ -0.00_{\pm0.01}$ & $ -0.03_{\pm0.01}$ & $ -0.07_{\pm0.01}$ & $ -0.13_{\pm0.01}$ & $ -0.19_{\pm0.01}$ & $ -0.25_{\pm0.01}$ & $ -0.32_{\pm0.01}$ & $ -0.41_{\pm0.01}$ & $ -0.48_{\pm0.01}$ \\
  & $[13.60, 14.00]$ & $ -0.11_{\pm0.02}$ & $ -0.08_{\pm0.01}$ & $ -0.07_{\pm0.01}$ & $ -0.09_{\pm0.01}$ & $ -0.15_{\pm0.01}$ & $ -0.22_{\pm0.01}$ & $ -0.27_{\pm0.01}$ & $ -0.31_{\pm0.01}$ & $ -0.34_{\pm0.01}$ & $ -0.39_{\pm0.01}$ & $ -0.46_{\pm0.01}$ & $ -0.54_{\pm0.01}$ \\
  & $[14.00, 14.50]$ & $ -0.21_{\pm0.02}$ & $ -0.20_{\pm0.02}$ & $ -0.20_{\pm0.02}$ & $ -0.25_{\pm0.02}$ & $ -0.32_{\pm0.02}$ & $ -0.38_{\pm0.02}$ & $ -0.41_{\pm0.02}$ & $ -0.41_{\pm0.02}$ & $ -0.40_{\pm0.02}$ & $ -0.40_{\pm0.02}$ & $ -0.48_{\pm0.02}$ & $ -0.56_{\pm0.01}$ \\
\hline
\end{tabular}
}
\end{sidewaystable}

\end{document}